\documentclass[twoside,twocolumn]{article}

\RequirePackage{inputenc,crop,graphicx,amsmath,array,color,amssymb,flushend,stfloats,amsthm,chngpage,times,verbatim}
\RequirePackage[LY1]{fontenc}

\def\scriptsize{\@setfontsize\scriptsize{6.5pt}{9.5pt}}
\def\scriptsize{\@setfontsize\normalsize{8}{11}}

\usepackage[table]{xcolor}
\definecolor{matlab}{RGB}{255,60,0}
\definecolor{R}{RGB}{60,170,240}

\usepackage[english]{babel} 
\usepackage{booktabs} 

\usepackage{fancyhdr} 
\usepackage{titling} 

\usepackage{hyperref} 

\addto\captionsenglish{}

\pretitle{\begin{center}\Huge\bfseries} 
\posttitle{\end{center}} 

\title{Rcall: Calling R from MATLAB}
\author{
\textsc{Janine Egert$^{\sf 1,2}$}\thanks{Corresponding author, \href{mailto:egert@imbi.uni-freiburg.de}{egert@imbi.uni-freiburg.de} }
\and 
\textsc{Clemens Kreutz$^{\sf 1,2,3}$} 
}
\date{$^{\text{\sf 1}}$ Institute of Medical Biometry and Statistics (IMBI), Faculty of Medicine and Medical Center Freiburg - University of Freiburg, 79104 Freiburg, Germany \\
$^{\text{\sf 2}}$ Centre for Integrative Biological Signalling Studies (CIBSS), University of Freiburg, 79104 Freiburg, Germany\\
$^{\text{\sf 3}}$ Center for Data Analysis and Modelling (FDM), University of Freiburg, 79104 Freiburg, Germany} 
\renewcommand{\maketitlehookd}{%
\vspace{-15pt}
\begin{abstract}
\noindent
\textbf{Summary:}
R and Matlab are two high-level scientific programming languages which are frequently applied in computational biology. To extend the wide variety of available and approved implementations, we present the Rcall interface which runs in MATLAB and provides direct access to methods and software packages implemented in R. Rcall involves passing the relevant data to R, executing the specified R commands and forwarding the results to MATLAB for further use. The evaluation and conversion of the basic data types in R and MATLAB are provided.
Due to the easy embedding of R facilities, Rcall greatly extends the functionality of the MATLAB programming language.\\
\textbf{Availability:} Source code is freely available at
https://github.com/kreutz-lab/Rcall, implemented in MATLAB and supported on Linux, MS Windows and Mac OS X.
\end{abstract}
}

\begin{document}
\maketitle

\section{Introduction}

With the computational and methodological advances in the field of bioinformatics, more and more software tools and functionalities have been developed in the last decades.
However, many algorithms are available only for one programming language, which prevents rapid implementation of state-of-the-art methods. Translating existing tools or switching software is undesirable and time consuming. For this reason, there is a need for function interfaces that enable evaluations of functions implemented in another programming language.

MATLAB is one of the most commonly used software in disciplines such as engineering and physical science. R is one of the most frequently applied programs for statistical computations. An R interface for MATLAB combines the full feature set of two high-level scientific computing software. Especially due to the large community effort in R, an R interface greatly augments the MATLAB programming language with existing well-tested and approved algorithms from the R community.

Existing R interfaces for the MATLAB environment are restricted to specific operating systems, e.g. the package 'RMatlab' (\cite{Lang2004}) for Unix systems, 
the R server 'Rserve' (\cite{Urbanek}) for Unix and Mac OS systems,
and the 'MATLAB R-link' (\cite{Henson2021}) for Windows systems. 
The 'MATLAB R-link' is based on the 'R (D)COM' server (\cite{Baier2008}). However, the non-commercial version of the 'R (D)COM' server on the CRAN archive is deprecated. 

We present the Rcall interface, which allows the direct application of R packages within the MATLAB environment.
Other R function interfaces are more flexible and allow calling R from
multiple programming languages (’MATLAB R-link’) or also support bidirectional functionality (’RMatlab’).
Rcall aims to extend the available methods for communicating with R
by offering a well-functioning and easy-to-integrate interface for all common operating systems, in particular also Windows systems are supported.


\section{Implementation}

\begin{figure*}[t]
\centering\includegraphics[width=\textwidth]{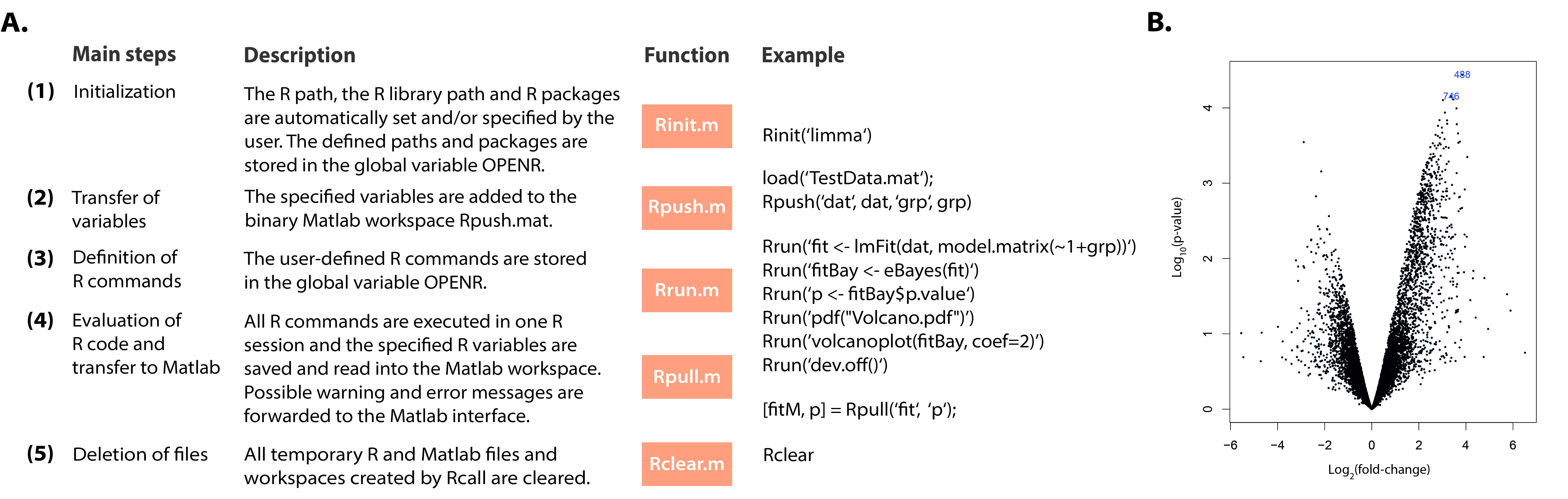}
\caption{\textbf{A.} The main steps of the Rcall implementation are described and illustrated. The main functions of the Rcall interface are highlighted by a light orange box. Usage of the R package 'limma' is shown as an example. \textbf{B.} The volcano plot of the exemplary differential expression analysis. The top two genes are highlighted by their gene ID.}\label{Fig}
\end{figure*}

Once the Rcall folder is added to the MATLAB path, Rcall is available in any MATLAB script without any installation procedure. The only requirements for Rcall are running MATLAB and R installations including the applied R packages. 

The implementation consists of five main steps which are illustrated in Figure \ref{Fig}.
First, Rcall is initialized (1). The R libraries, R path, and the path to the R libraries can be specified by the user, otherwise the current system configurations are used automatically. In Windows, the latest R version on the system is located.
The variables defined in the \textit{Rpush} command are stored in a temporary MATLAB workspace file by using the `R.Matlab' package (\cite{Bengtsson2018}) (2). With each call of the \textit{Rpush} command the temporary MATLAB workspace file is expanded. 

The R commands are specified in MATLAB by the \textit{Rrun} function as string argument (3). To reduce the amount of computation required to invoke a new R instance, the R commands are cached until a result is demanded via the \textit{Rpull} function. In the \textit{Rpull} function, the cached R commands are passed to an automatically generated, temporary R script. The R script is executed in batch mode via a system command line call (4). To reduce computation time, only R commands defined between two Rpull calls are evaluated. 
The input arguments of the \textit{Rpull} function specify which output variables are passed to a result MATLAB workspace file by the 'R.Matlab' package. All data types which are supported by the 'R.Matlab' package such as scalar, vector, matrix, array, character, etc. can be transferred.
To extend the functionality of the 'R.Matlab' package, Rcall assigns corresponding data types such as R data.frames to MATLAB tables and R lists to MATLAB structures, vice versa in the \textit{Rpush} command. An overview of the data type conversion is depicted in Table \ref{Tab}.
\begin{table}[t]
{\centering\begin{tabular}{@{} >{\columncolor{matlab!20}}l >{\columncolor{R!30}}l @{}}\toprule
MATLAB &R \\\midrule
Numeric & Numeric \\
Character & Character\\
Cell & List \\
Structure & List \\
Table & Data frame \\
\end{tabular}
\caption{Data type conversion implemented in Rcall.}\label{Tab}}{}
\end{table}
Nested object and list export is supported by passing the object elements individually to MATLAB and then combining them into a MATLAB structure.
In general, and especially for large or complex objects, it is more efficient to extract relevant information from the R object and pass only the extracted results to MATLAB.
To facilitate error handling, error messages that appear in R are forwarded to the MATLAB console. All scripts and variables remain stored until they are cleared by the \textit{Rclear} function (5). This allows repeated use of the variables and objects.

To illustrate the use of Rcall, a differential expression analysis is performed on a simulated gene expression data set using the R package 'limma' (\cite{limma}).
The data $dat$ consists of 10,000 log$_2$-normally distributed genes which consists of two probes with three replicates each. In the second probe, 10\,\% of the genes are differentially expressed. The predictor vector $grp$ corresponds to the sample indices. The \textit{lmFit} R function fits a linear model to the gene expression data. The empirical Bayes adjusted t-statistics and p-values are calculated by the \textit{eBayes} R function. To visualize the differential expression, a volcano plot is generated and saved using the \textit{pdf} R function (Figure \ref{Fig}B.).

The results of the linear model fit are stored in a MArrayLM object \textit{fit }which is converted to a structure \textit{fitM} in MATLAB. Alternatively, the object elements can be assigned to separate variables and transferred individually to MATLAB which is shown in the example for the p-values \textit{p}.
The execution of the example code took on average 4.5 seconds real time in MATLAB and 0.8 seconds in R on a 2.60GHz Intel(R) Core(TM) i5-7300U with 4 logical processors. 

Rcall invokes any R functionality from object creation to code evaluation and visualization. Major limitations of Rcall are that R commands requiring user input are not supported, complex data objects can only be transferred if conversion to list is possible, and rendering of graphics is restricted to functionality that is provided by R in batch mode.


\section{Conclusion}

Rcall enables the direct execution of R commands from the MATLAB work environment.
Evaluation and transfer of basic data types is supported. The variables are converted to the corresponding data types in R and MATLAB.
Rcall provides an easy-to-integrate system and due to its execution in batch mode, it works on all major operating systems (Unix, MS Windows, Mac OS X) without sophisticated installation or configuration procedures. No special settings or third party software are required.

\section{Funding}
This work has been supported by the Federal Ministry of Education and Research of Germany [EA:Sys,FKZ031L0080 to J.E. and C.K.]; and the Excellence Initiative of the German Federal and State Governments [CIBSS-EXC-2189-2100249960-390939984 to C.K.].

\section{Conflict of Interest}
The authors declare that there is no conflict of interest regarding the publication of this article.

\bibliographystyle{apalike}
\bibliography{Rcall_Egert}

\begin{thebibliography}{}

\bibitem[Baier, 2008]{Baier2008}
Baier, T. (2008).
\newblock {\em {R/Scilab (D)Com Server 3.0-1B5}}.
\newblock https://cran.r-project.org/contrib/extra/dcom/.

\bibitem[Bengtsson and Jacobson, 2018]{Bengtsson2018}
Bengtsson, H. and Jacobson, A. (2018).
\newblock {\em {R.matlab: Read and Write MAT Files and Call MATLAB from Within
  R}}.
\newblock https://CRAN.R-project.org/package=R.matlab.

\bibitem[Henson, 2021]{Henson2021}
Henson, R. (2021).
\newblock {\em {MATLAB R-link}}.
\newblock {MATLAB Central File Exchange}.
\newblock https://www.mathworks.com/matlabcentral/fileexchange/
  5051-matlab-r-link.

\bibitem[Lang and Liu, 2004]{Lang2004}
Lang, D.~T. and Liu, B. (2004).
\newblock {\em RMatlab (Version 0.2-5)}.
\newblock http://www.omegahat.net/RMatlab/.

\bibitem[Ritchie et~al., 2015]{limma}
Ritchie, M.~E., Phipson, B., Wu, D., Hu, Y., Law, C.~W., Shi, W., and Smyth,
  G.~K. (2015).
\newblock {limma} powers differential expression analyses for {RNA}-sequencing
  and microarray studies.
\newblock {\em Nucleic Acids Research}, 43(7):e47.

\bibitem[Urbanek, 2019]{Urbanek}
Urbanek, S. (2019).
\newblock {\em {Rserve: Binary R server}}.
\newblock https://CRAN.R-project.org/package=Rserve.

\end{thebibliography}


\end{document}